\documentclass[aps,superscriptaddress,groupedaddress,twocolumn,showkeys,showpacs,byrevtex]{revtex4}

\usepackage{graphicx}
\begin{document}

\title{Experimental verification of energy correlations in entangled photon pairs}

\author{Jan Soubusta}\email{soubusta@sloup.upol.cz}
\author{Jan Pe\v{r}ina Jr.}
\author{Martin Hendrych}
\author{Ond\v{r}ej Haderka}
\affiliation{Joint Laboratory of Optics of Palack\'{y} University and
Institute of Physics of Academy of Sciences of the Czech Republic, 17. listopadu
50A, 772 00 Olomouc, Czech Republic}
\affiliation{Department of Optics, Palack\'{y}
University, 17. listopadu 50, 772 00 Olomouc, Czech Republic}
\author{Pavel Trojek}
\author{Miloslav Du\v{s}ek}
\affiliation{Department of Optics, Palack\'{y}
University, 17. listopadu 50, 772 00 Olomouc, Czech Republic}

\date{\today}

\begin{abstract}
Properties of entangled photon pairs generated in spontaneous parametric
down-conversion are investigated in interference experiments. Strong
energy correlations are demonstrated in a direct way. If a signal photon
is detected behind a narrow spectral filter, then interference appears
in the Mach-Zehnder interferometer placed in the route of the idler photon,
even if the path difference in the interferometer exceeds the coherence
length of the light. Narrow time correlations of the detection instants
are demonstrated for the same photon-pair source using the Hong-Ou-Mandel
interferometer. Both these two effects may be exhibited only by an
entangled state.
\end{abstract}

\maketitle


\section{Introduction}

Spontaneous parametric down-conversion (SPDC) \cite{Hong_PRA31} is
a widespread method used for generation of entangled pairs of
photons. These highly correlated particles are favourable for
testing many interesting features of quantum mechanics
\cite{PerinaHradil}.

Transverse correlations in SPDC were theoretically analyzed in
detail by Rubin \cite{Rubin_PRA54}. Transfer of angular spectrum
between SPDC beams was used for optical imaging by means of
two-photon optics \cite{Pittman_PRA52,Pittman_PRA53}. The angular
spectrum is also transferred from the pump beam to both
down-converted beams, as examined, e.g., in Ref.~\cite{Monken_PRA57}.
The coherence times of the down-converted photons were measured
first in the 80's by Hong, Ou and Mandel (HOM) \cite{Hong_PRL59}.
The HOM interferometer was used in many experiments testing 
interference from the indistinguishability point of view
\cite{Pittman_PRL77,Kwiat_PRA45}. 
Placing linear polarizers into both HOM interferometer arms enables
demonstration of violation of Bell's inequalities, measuring
correlations of mixed signal and idler photons from the pair as
a function of the two polarizer settings \cite{Ou_PRL61}.

Franson \cite{Franson_PRL62} suggested another experiment. By use of two 
spatially separated interferometers, one can observe a violation of Bell's
inequalities for position and time. First high-visibility interference 
experiments of this type were performed using Michelson interferometer 
\cite{Brendel_PRL66}, reaching a visibility of 87\%. Later experiments 
utilized spatially separated Mach-Zehnder interferometers to measure nonlocal 
interference \cite{Rarity_PRA45,Kwiat_PRA47}.
Interference was observed also in mixing of signal photons produced 
by SPDC in two nonlinear crystals \cite{Zou_PRL69}. 
It was demonstrated that when the optical path difference through this 
interferometer exceeds the coherence length, interference appears 
as a spectrum modulation. Finally, complementarity between
interference and a which-path measurement was studied for the case
of down-converted light in Ref. \cite{Bjork}.

The aim of this work is to study the correlations of energies of
the two photons from the entangled pair directly. To achieve this
goal, we place a Mach-Zehnder (MZ) interferometer in the idler
beam and set the optical-path difference larger than the
coherence length of the idler photons. Thus, no
interference pattern is observed upon modulation of the 
optical-path difference on a scale of the wavelength. If a narrow
frequency filter were placed in the idler beam in front of the MZ
interferometer, the coherence length would be effectively prolonged 
and the interference pattern would be observed.
Similarly, placing a scanning filter in front of one detector at
the output of the MZ interferometer, one could observe an intensity
modulation of particular frequency component,
because individual frequency components of the field
interfere independently and the filter selects just one
quasi-monochromatic portion of them.
However, there is no filter in the idler-beam path containing the 
MZ interferometer in the setup described in this paper. We place the
narrow frequency filter in the signal beam and observe an effective 
prolongation of the coherence length in the MZ interferometer and 
a substantial increase of visibility in a coincidence-count measurement.
This prolongation is the consequence of the correlation of energies of 
the two beams \cite{Kwiat_PRL66,Dusek_CJP46,Trojek_CJP}.
Kwiat and Chiao used similar setup in their pioneering work in experiments 
testing Berry's phase at the single-photon level \cite{Kwiat_PRL66}. 
By filtering the signal photon by a "remote" 
filter with a bandwidth of 0.86~nm, they observed in coincidence 
measurement the interference fringes visibility of 60~\% 
behind a Michelson interferometer detuned by 220~$\mu$m. 
They interpreted these results in terms of a nonlocal collapse 
of the wave function. In this paper we extend these original results 
by utilizing a one-order-of-magnitude narrower frequency filter, the 
Fabry-Perot (FP) resonator. Moreover, the visibility dependence 
on the MZ interferometer detuning is derived and measured.

The paper is organized as follows. In section \ref{sec_experiment}, we
describe the experimental setup. In section \ref{sec_theory}, we give
the basic formulas describing this experimental situation. In section
\ref{sec_energy}, the energy correlations are measured experimentally
and the agreement with the theory is verified. It is important to
note that the increase of visibility described above could also be
observed with optical fields exhibiting classical correlations 
\cite{Kwiat_PRL66,Dusek_CJP46}. 
To exclude this possibility we also test time correlations of the photon 
pairs generated by our source \cite{Kwiat_PRA47,Franson_PRL67}.
This is shown and discussed in section \ref{sec_time}.
Section \ref{sec_conclusions} concludes the paper.


\section{Experiment}\label{sec_experiment}

\begin{figure}[tbh]
\centerline{\includegraphics[width=\columnwidth]{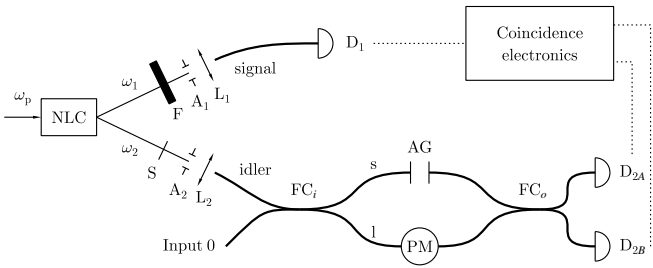}}
\caption{Experimental setup (see text for details).}\label{setup}
\end{figure}

A scheme of the experimental setup is shown in Fig.~\ref{setup}.
Type-I parametric down-conversion occurs in a LiIO$_3$ nonlinear
crystal (NLC) pumped by a krypton-ion cw laser at frequency
$\omega_p$ (413.1~nm). Signal photons $\omega_1$ and idler
photons $\omega_2$ ($\omega_1+\omega_2=\omega_p$) are selected by
apertures A$_1$ and A$_2$, respectively, and coupled into fibers 
by lenses $L_1$ and $L_2$.
A frequency filter F is placed in the signal-beam path. S is a
shutter optionally closing the idler beam. AG is an air-gap that
allows to set the length of one arm (called s) of the fiber-based MZ
interferometer with a precision of 0.1~$\mu$m. Phase modulator PM
is used to change the optical path of the other arm (called l) of
the interferometer on a wavelength scale to observe the interference
pattern. FC$_i$ and FC$_o$ denote input and output fiber couplers 
respectively. D$_1$, D$_{2A}$, D$_{2B}$ denote Perkin-Elmer SPCM
photodetectors. Detection electronics record simultaneous
detections (within a 4~ns time window) of detectors D$_1$ and
D$_{2A}$, respectively of detectors D$_1$ and D$_{2B}$. These data are
transferred to a PC as a function of the air-gap position ($l_{AG}$)
and the phase set by the PM.

Because of thermal fluctuations, the difference of optical lengths
of the fiber MZ interferometer arms drifts on the scale of
several nanometers per second. Therefore it is necessary to actively
stabilize the interferometer. We use attenuated pulsed
laser-diode beam connected to the free input of the interferometer
(Input 0 in Fig.~\ref{setup}) for the stabilization. The
measurements run in cycles repeating two steps. First, the
down-conversion idler beam is blocked by the shutter S, and the
pulsed laser is switched on. A zero phase position of the phase modulator 
is found that corresponds to minimum counts on the detector D$_{2A}$.
Next, the pulsed laser is switched off and the shutter is opened.
The phase in the MZ interferometer is then set
against the last found zero phase position and the measurement is
performed. A time duration of one measurement cycle is 10 seconds.

We perform four different measurements distinguished by the use
of different filters. First, we use the setup described in 
Fig.~\ref{setup} without any filter in the signal-photon path. The
visibility of the coincidence-count interference pattern is peaked 
around the balanced MZ interferometer with a FWHM of 160~$\mu$m. 
Second, we insert a narrow band interference filter in the 
signal-photon path and the visibility peak broadens (350~$\mu$m). 
Third, we use a FP resonator to filter the signal 
photons. Then the high-visibility region is broadened considerably,
but the measured dependence is modulated with a period
corresponding to one roundtrip in the FP resonator. Finally, 
we insert both filters (the FP resonator and the narrow band interference
filter) in the signal-photon path. In this case no oscillations appear
in the measured visibility of the coincidence-count interference. 
The theory presented in the next section is intended to explain 
this behavior of the visibility pattern for the above specified cases.


\section{Theory}\label{sec_theory}

In this section we briefly mention theoretical description of the behavior
of photon pairs generated by SPDC in our experimental setup (Fig.~\ref{setup})
and define the used notation.
The theoretical coincidence-count interference visibility is calculated
for the Gaussian filter (section \ref{subsec_gauss}) and for the Gaussian filter
together with the FP resonator (section \ref{subsec_FP}).

SPDC is commonly
studied in the first-order perturbation theory which gives for
the cw pumping at the carrying frequency $ \omega_p $ a two-photon
state $ |\psi^{(1)}\rangle $ at the output plane of the nonlinear
crystal \cite{Ou_PRA40,Grice_PRA56,Perina_PRA59,Perina_EPJD7,Atature_PRA66,Nambu_PRA66}
\begin{widetext}
\begin{equation}\label{Psi2}
 | \psi ^{(1)} \rangle = C_{\psi} \int d\omega_1 \int d\omega_2
 \Phi(\omega_1,\omega_2) \delta(\omega_1+\omega_2-\omega_p)
 \hat{a}^{\dagger}_1 (\omega_1) \hat{a}^{\dagger}_2 (\omega_2)
 | {\rm vac} \rangle.
\end{equation}
\end{widetext}
The symbol $\hat{a}^{\dagger}_j (\omega_j)$, $j=1,2,$ stands
for the creation operator of a photon with the frequency
$\omega_{j}$ in the field $ j $. The $\delta$ function in
Eq.~(\ref{Psi2}) expresses the energy conservation law. The
phase-matching function $\Phi(\omega_1,\omega_2)$ describes
correlations between modes in the signal and idler fields and is
found in the form
\begin{equation} \Phi(\omega_1,\omega_2)=\frac
{1-\exp[i(\omega_1+\omega_2-\omega_p)L/v_{g}]}
{-i(\omega_1+\omega_2-\omega_p)/v_{g}},
\end{equation}
$L$ is the length of
the nonlinear crystal, $v_g$ is the group velocity of the
down-converted light (identical for signal and idler photons in
degenerate type-I process).

Photons generated via SPDC are conveniently described with the use of a
two-photon probability amplitude ${A}$ defined as \begin{equation}
\label{A}
 {A} (\tau_{1},\tau_{2}) = \langle {\rm vac} |
 \hat{E}^{(+)}_{1} (\tau_{1}) \hat{E}^{(+)}_{2}(\tau_{2})
 | \psi ^{(1)} \rangle,
\end{equation}
which provides the two-photon amplitude behind the apertures. The symbol
$\hat{E}^{(+)}_{j}$  denotes the positive-frequency part of the electric
field operator of the $j$-th down-converted field. The two-photon
amplitude ${A}_{MZI}$ appropriate for the case when both photons are
located just at the detectors D$_1$ and D$_{2A}$, or D$_1$ and D$_{2B}$
(see Fig.~\ref{setup}) can be expressed in the form \cite{Trojek_CJP}
\begin{equation} \label{Amzi}
 {A}_{MZI}(\tau_1,\tau_2) = {T}_s {A}(\tau_1,
 \tau_2) + {T}_l{A}(\tau_1,\tau_2 + \Delta t),
\end{equation}
i.e., as a sum of two contributions, each representing one
alternative path of the idler-photon propagation through the MZ
interferometer. Here, $\Delta t$ is the propagation time difference 
of the two interferometer arms, ${T}_s$ and ${T}_l$ are
the overall amplitude transmittances of the two alternative
paths.

The coincidence-count rate
$R_c$ measured in this experiment can be represented in terms
of the two-photon amplitude ${A}_{MZI}$ as follows
\begin{equation} \label{Rc}
 R_c (\Delta t) = \int_{- \infty}^{\infty} d\tau_1
 \int_{- \infty}^{\infty} d\tau_2 \left| {A}_{MZI} (\tau_1,\tau_2)
 \right| ^2.
\end{equation}
The normalized coincidence-count rate $R_n$ then reads
\begin{equation} \label{Rn}
 R_n (\Delta t) = 1 + \rho (\Delta t),
\end{equation}
where $\rho$ is an interference term:
\begin{widetext}
\begin{equation}
 \rho(\Delta t) = \frac{ 2{\rm Re}\left\{ {T}_s
 {T}^*_l
 \int_{-\infty}^{\infty}
 d\tau_1  \int_{-\infty}^{\infty} d\tau_2
 {A}(\tau_1,\tau_2) {A}^*(\tau_1,\tau_2+ \Delta t)
 \right\}  }{
 (|{T}_s|^2 + |{T}_l|^2)
 \int_{-\infty}^{\infty}
 d\tau_1  \int_{-\infty}^{\infty} d\tau_2
 |{A}(\tau_1,\tau_2)|^2 }.
\end{equation}
\end{widetext}
The symbol $ {\rm Re} $ denotes the real part of the argument.
The detuning $ \Delta t $ is given as $\Delta t = l_{AG}/c$,
where $l_{AG}$ is the air-gap position ($l_{AG}=0$ corresponds to
the balanced interferometer) and $c$ is the speed of light.
The visibility $V$ of the coincidence-count interference pattern
described by $ R_n(\Delta t) $ for a given detuning $\Delta t$ of the MZ
interferometer is calculated according to
\begin{equation}
V = \frac {R_{n,\rm max}-R_{n,\rm min}}{R_{n,\rm max}+R_{n,\rm min}}.
\end{equation}
The quantities $R_{n,\rm max}$ and $R_{n,\rm min}$
are the maximum and minimum values obtained in one optical period of
the interference pattern.


\subsection{Setup with a Gaussian filter}\label{subsec_gauss}
In real experiments, the idler and signal beams are usually filtered
by interference filters before they are coupled to the fibers to minimize
the noise counts arising from stray light in the laboratory. Even in
the absence of interference filters (cut-off filters blocking the
scattered UV pump and luminescence are 
sufficient), geometric frequency filtering occurs due to a limited aperture
of the fiber-coupling optics. This is due to the fact that SPDC
output directions of the phase-matched photons of a pair depend on
a particular combination of their frequencies  $ \omega_1, \omega_2 $.

The influence of the interference filters and the filtering due to the 
coupling apertures can be approximately modelled using Gaussian filters
with the intensity transmittance given by the expression
\begin{equation} \label{T_g}
  T_{Gj}(\omega_j) = \exp \left[ -\frac {(\omega_j-\omega_j^0)^2}
                                        {\sigma_j^2}
                          \right], \quad j=1,2.
\end{equation}
$\omega_j^0$ is the central frequency and the FWHM of the filter 
intensity profile obtainable experimentally is given by 
$2\sqrt{\ln 2}\,\sigma_j$.
In the case without additional filters, $\sigma_1=\sigma_2$.

The interference term $\rho$ is given by
\begin{eqnarray}
&&\rho (\Delta t) = \frac {2
 {\rm Re} \left\{{T}_l^* {T}_s \exp \left[i\omega_2^0\Delta t\right]\right\}}
 {|{T}_l|^2 + |{T}_s|^2}
 \exp \left[-\frac{\Delta t^2}{4\beta_2}\right], \nonumber \\
&&\mbox{where}\quad
\beta_2 = \frac 1{\sigma_1^2} + \frac 1{\sigma_2^2}.\label{vis_gauss}
\end{eqnarray}
In the ideal case, when both paths in the MZ interferometer are
equally probable (${T}_l={T}_s$), the interference term
can be simplified to the form
\begin{equation}
\rho (\Delta t) = \cos (\omega_2^0\Delta t) \cdot
   \exp \left[-\frac{\Delta t^2}{4\beta_2}\right].
\end{equation}
The first function (cosine) represents a
rapidly oscillating function, while the other function (Gaussian)
describes the envelope of the oscillations. The local visibility
of the coincidence-count interference pattern $ R_n $ around a certain
value $\Delta t$ of the interferometer detuning is then equal to
the envelope function
\begin{equation} \label{vis_g}
  V (\Delta t) = \exp \left[-\frac{\Delta t^2}{4\beta_2}\right].
\end{equation}


\subsection{Setup with a Fabry-Perot resonator}\label{subsec_FP}
Next, we use a FP resonator as a tunable filter in the
signal-photon path. This resonator can provide a two-orders-of-magnitude 
narrower frequency filter in comparison with the above
used Gaussian filter. It should be stressed that the geometric filtering
mentioned above also takes place in this case. The intensity transmittance
of the FP resonator is given by the expression \cite{SalehTeich}
\begin{equation} \label{T_FP}
T_{FP1}(\omega_1) = \frac {T_{\rm max}} {1 + \left(\frac {2F}{\pi}\right)^2
                    \sin^2 \left( \frac {l_F}{c}\omega_1 \right)},
\end{equation}
which is a periodic function with narrow peaks of maximum
transmittance $T_{\rm max}$ at a distance given by free spectral
range $\omega_{FSR}=\pi c/l_F$. Parameter $F$ is the finesse of
the resonator. The width of the transmittance peaks is inversely
proportional to the finesse.

In this case the interference term is found to be
\begin{widetext}
\begin{equation}
\rho (\Delta t) = \frac {2{\rm Re} \left\{{T}_l^* {T}_s \exp \left[
      i\omega_2^0\Delta t \right]
                  \int\limits_{-\infty}^{\infty} d\nu_1
                  \frac {\exp \left[-\nu_1^2\beta_2-i\nu_1\Delta t \right]}
                        {1+\left(\frac {2F}{\pi}\right)^2 \sin^2
                               \left(\frac {l_F}{c}(\omega_1^0+\nu_1)\right)}
           \right\}}
      {(|{T}_l|^2 + |{T}_s|^2)\int\limits_{-\infty}^{\infty} d\nu_1
            \frac {\exp \left[-\nu_1^2\beta_2 \right]}
                  {1+\left(\frac {2F}{\pi}\right)^2 \sin^2
                               \left(\frac {l_F}{c}(\omega_1^0+\nu_1)\right)}}.
\end{equation}
Performing the integration we can introduce a new expression
for the interference term
\begin{equation}
 \rho (\Delta t) = \frac 2C {\rm Re} \left\{ {T}_l^* {T}_s
   \sum\limits_{n=-\infty}^{\infty}
     \left(1 + \frac 2{\gamma} - \frac 2{\gamma} \sqrt{1+\gamma} \right)^{|n|}
     \exp\left[i(\omega_2^0\Delta t + n \varphi_0)\right]
     \exp\left[-\frac{\left( \Delta t - nt_0\right)^2}{4\beta_2}\right]
  \right\}.
\label{vis_fp}
\end{equation}
The new parameters $C, t_0, \varphi_0 $, and $ \gamma$ are defined
as follows
\begin{eqnarray}
C &=& (|{T}_l|^2 + |{T}_s|^2)
    \left(1 + 2\sum\limits_{n=1}^{\infty}
      \left(1 + \frac 2{\gamma} - \frac 2{\gamma} \sqrt{1+\gamma} \right)^n
      \cos (n\varphi_0)
      \exp\left[ -\frac {n^2 t_0^2}{4\beta_2} \right]
    \right), \nonumber \\
 t_0 &=& 2 \frac {l_F}{c},  \quad
 \varphi_0 = 2\frac {l_F}{c} \omega_1^0, \quad \mbox{and}\quad
 \gamma   = \left( \frac {2F}{\pi} \right)^2.
\end{eqnarray}
\end{widetext}

The symbol $t_0$ determines the time of one roundtrip in the FP resonator,
$\varphi_0 = \omega_1^0 t_0$ is a phase factor acquired in one roundtrip.
The $n$-th term of the sum determining $ \rho $ in Eq.~(\ref{vis_fp})
describes the case in which the signal photon
propagates just $|n|$ times through the FP resonator and then
leaves the resonator. Owing to entanglement of photons in a pair,
the time structure of signal field introduced by the FP resonator
\cite{Perina_FP} is reflected in the idler-field interference
pattern in the MZ interferometer.
The $n$-th term contribution is peaked around the detuning value
$\Delta t = n t_0$. The maximal value reached for this detuning equals
$ \left( 1 + 2/\gamma - 2\sqrt{1+\gamma} / \gamma \right)^{|n|} $.
The width of the peak is the same as in the case without
the FP resonator, it is determined by the
spectral width of the entangled photons. If the width of the peak
is smaller than $t_0$, then the visibility resulting from (\ref{vis_fp})
is a modulated function. We can get the following approximate expression
for the upper envelope of this visibility dependence:
\begin{equation}
 V^{\rm upper}(\Delta t) = \left( 1 + \frac{2}{\gamma}
 -  \frac{2\sqrt{1+\gamma} }{\gamma}  \right)^{ \frac{|\Delta t|} {t_0} } .
 \label{upper}
\end{equation}
We note that the approximate expression for $V^{\rm upper}$
in Eq.~(\ref{upper}) may be used even if the width of peaks
is greater than $t_0$.

One can also compare the results given in Eqs. (\ref{vis_fp}) and (\ref{vis_gauss})
from another point of view. In the setup without the FP resonator, detection of the
signal photon by detector D$_1$ yields the time of the pair creation. Therefore,
one can distinguish, at least in principle, the path of the idler
photon through the unbalanced MZ interferometer from the detection
time at the detector D$_{2A}$. Hence no interference can be observed.
If the MZ interferometer's detuning is equal to $n$ roundtrips in the FP
resonator which is placed in the signal-photon path, one can not distinguish 
between two possible events: a) either the idler photon takes the shorter arm 
of the MZ and the signal photon makes $m$ roundtrips in the FP, where $m$ is an 
integer; b) or the idler photon takes the longer arm of the MZ and the signal 
photon makes $m+n$ roundtrips. 
This option of two indistinguishable paths restores the interference 
\cite{Franson_PRL62,Brendel_PRL66,Rarity_PRA45,Kwiat_PRA47}.
The longer detuning $ \Delta t $, the more unbalanced 
interfering probability amplitudes (corresponding to cases a) and
b)) and consequently the lower visibility.


\section{Energy correlations}\label{sec_energy}

\begin{figure}[hbt]
\centerline{\includegraphics[height=\columnwidth,angle=270]{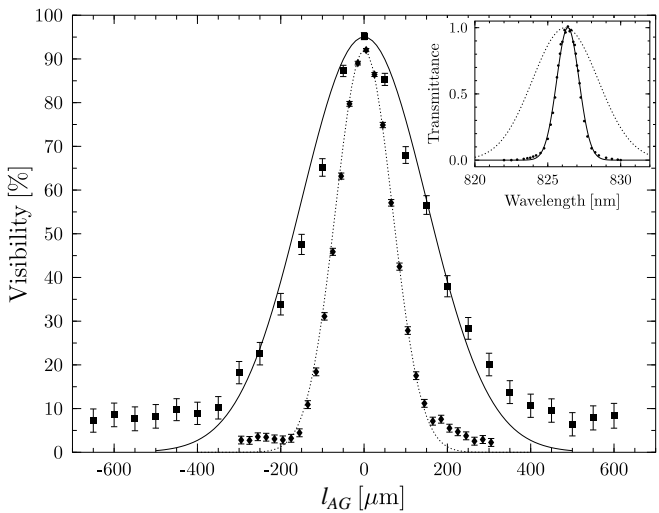}}
\caption{Coincidence-count interference visibility between
detectors D$_{2A}$ and D$_1$ as a function of the air-gap
position. Diamonds correspond to the measurement without any
filter. The dotted curve is a Gaussian fit according to
Eq.~(\protect\ref{vis_g}) yielding the geometric spectral 
filtering of FWHM=5.3~nm. The spectrum calculated according to 
Eq.~(\protect\ref{T_g}) is plotted with the dotted curve 
in the inset. Squares correspond to the measurement
with the narrow band interference filter. The continuous curve is
the theoretical dependence obtained for the parameters of the filter
provided by the manufacturer.
The inset shows the normalized transmittance of the filter (dots) and
a Gaussian fit of these data (continuous curve) yielding a
FWHM of 1.8~nm.} \label{Fig_Gauss}
\end{figure}

First we use the setup shown in Fig.~\ref{setup} without the FP resonator.
Figure~\ref{Fig_Gauss} shows the visibility of interference
pattern in the measured coincidence-count rate between detectors
D$_{2A}$ and D$_1$ as a function of the air-gap position. The
visibility reaches significant values only in a region whose width
is inversely proportional to the bandwidths of the signal and
idler spectra (see Eq.~(\ref{vis_g})). Diamonds correspond to the
measurement without any filter, thus showing the effect of the 
geometric filtering of the fiber-coupling optics. 
Plotted errors represent the standard deviations. 
The dotted curve is a Gaussian fit of these data with a FWHM of 160~$\mu$m,
which corresponds to a spectrum of 5.3~nm FWHM. This spectrum of 
the signal photons, calculated according to Eq.~(\protect\ref{T_g}), 
is plotted with the dotted curve in the inset of Fig.~\ref{Fig_Gauss}.

Squares in Fig.~\ref{Fig_Gauss} correspond to the coincidence-count 
interference visibility measured in the setup modified by inserting the 
narrow band interference filter centered at 826.4~nm (Andover Corporation) 
in the signal-beam path. Dots in the inset show the measured normalized 
transmittance of the filter provided by the manufacturer, and the continuous 
curve is a Gaussian fit of these experimental data with a FWHM of 1.8~nm.
The parameters of the filter yield
theoretical visibility dependence of a 350~$\mu$m
FWHM (continuous curve). This curve fits well the
experimental data (squares) in the central part of the pattern.
The deviations from the Gaussian shape in the low-visibility wings
are caused by the difference of the real spectrum of the filters
from the Gaussian shape, as can be shown by a more detailed
analysis tractable only numerically.
The narrowing of the
spectrum of signal photons from 5.3 nm to 1.8 nm thus leads to the
broadening of the
coincidence-count interference pattern from 160~$\mu$m to 350~$\mu$m.

\begin{figure}[hbt]
\centerline{\includegraphics[height=\columnwidth,angle=270]{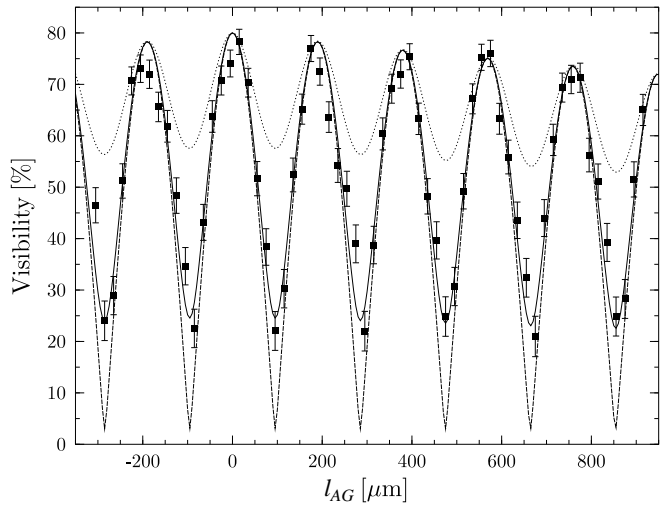}}
\caption{Coincidence-count interference visibility between
detectors D$_{2A}$ and D$_1$ as a function of the air-gap
position $ l_{AG} $. Squares correspond to the measurement setup with the
FP resonator in the signal-photon path. Curves are the
fits given according to Eq.~(\protect\ref{vis_fp}) with the parameter
$F=150$. The dotted curve:
$l_F$=95.00~$\mu$m, the continuous curve: $l_F$=94.86~$\mu$m, the
dashed curve: $l_F$=94.80~$\mu$m.} \label{Fig_FP}
\end{figure}

Figure~\ref{Fig_FP} shows the coincidence-count interference visibility
measured with the FP resonator. The length of the resonator $l_F$
can be tuned over a range of 1.5~$\mu$m. The high-visibility region is now
considerably wider than the one obtained in the
previous case, but we get a modulated curve described by Eq.~(\ref{vis_fp})
instead of a single-peak function.
The period of the measured visibility oscillations is given by one
roundtrip in the FP resonator 2$l_F$, yielding the
resonator length $l_F\approx 95~\mu$m.
We note that similar oscillations in interference patterns
were observed in \cite{Gisin}, where properties of
entangled photon pairs generated by a train of pump pulses were studied.

\begin{figure}[htb]
\centerline{\includegraphics[height=0.75\columnwidth,angle=270]{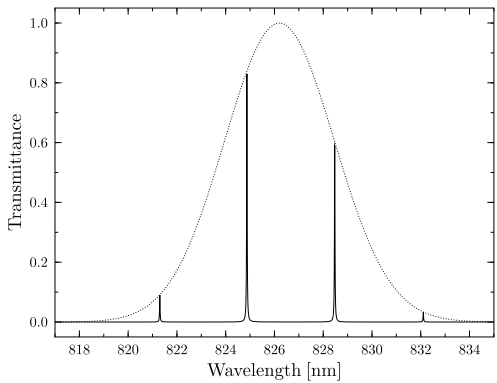}}
\caption{The continuous curve is the spectrum of the signal photons calculated
according to Eq.~(\protect\ref{T_FP}) that yields the measured 
visibility modulation of Fig.~\protect\ref{Fig_FP} ($l_F$=94.86~$\mu$m).
The dotted curve represents the geometric filtering
(FWHM=5.3~nm).}\label{Spek_FP}
\end{figure}

The free spectral range of the FP resonator expressed in units of wavelength
is $\lambda_{FSR} = \frac 12 \lambda^2/l_F \approx 3.6$~nm. Compared to
the 5.3~nm FWHM of the geometric spectrum of signal photons, it is
obvious that more than one peak of the FP contributes to the signal
at detector D$_1$. If we change the FP length $l_F$ by one 
quarter of the wavelength, the modulation of visibility changes 
from one limiting case to another according to the changes of the spectrum of 
the transmitted signal photons. $l_F=95.00~\mu$m
corresponds to the first limiting case. The theoretical spectrum has one dominant
peak at the center and two weak satellites; the visibility modulation is 
smallest (dotted curve in Fig.~\ref{Fig_FP}). For $l_F=94.80~\mu$m, the
visibility modulation is largest, corresponding to the second limiting case
(dashed curve in Fig.~\ref{Fig_FP}). In this case the spectrum is composed of
two dominant equivalently strong maxima. The other two small maxima prevent 
the visibility from reaching zero at its minima.
Comparing the theory with the experimental data, we have found that the FP
resonator length that provides the measured visibility modulation is 
$l_F=94.86\mu$m (solid line in Fig.~\ref{Fig_FP}). The spectrum corresponding
to this FP length is shown in Fig.~\ref{Spek_FP}.

\begin{figure}[thb]
\centerline{\includegraphics[height=\columnwidth,angle=270]{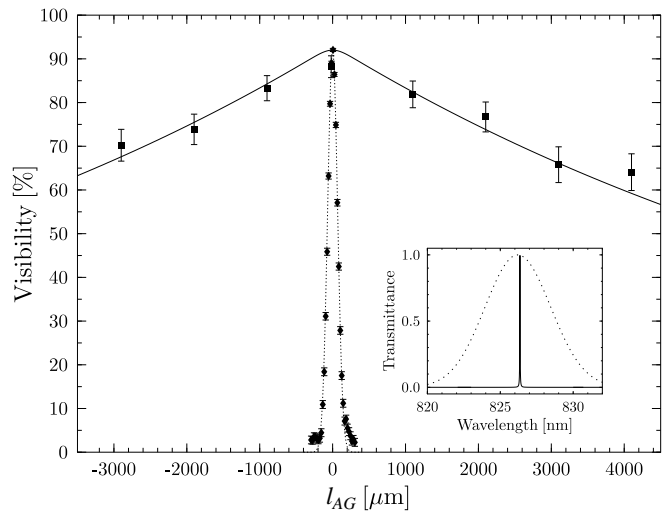}}
\caption{Coincidence-count interference visibility between
detectors D$_{2A}$ and D$_1$ as a function of the air-gap
position. Diamonds correspond to the measurement setup without any
filter. The dotted curve is a Gaussian fit given by
Eq.~(\protect\ref{vis_g}), corresponding to spectral FWHM=5.3~nm. Squares
correspond to the measurement setup with both the FP
resonator and the narrow band interference filter in the signal-photon
path. The continuous curve is the fit according to Eq.
(\protect\ref{vis_fp}) with parameters $F=150$ and
$l_F$=95.03~$\mu$m. The inset shows the calculated spectra of the
signal photons, Eq.~(\protect\ref{T_FP}). The dotted curve
corresponds to the geometric filtering (FWHM=5.3~nm), the continuous
curve represents the transmittance of the FP resonator together
with the narrow band filter.}\label{Fig_AFP}
\end{figure}

To avoid the effect of multiple frequency peaks transmitted
through the FP resonator, we use a combination of both filters
(the FP resonator and the narrow band interference filter)
in the signal-photon path. Figure~\ref{Fig_AFP} shows both the
measured coincidence-count interference visibility (squares) and
the theoretical curve (continuous curve) according to
Eq.~(\ref{vis_fp}). For comparison, the dependences obtained for
the setup with no filter are plotted as well (diamonds and
dotted curve, same as in Fig.~\ref{Fig_Gauss}). Using both filters
in the signal-photon path we are able to select just one narrow
transmittance peak of the FP resonator. This is
possible because the narrow band filter's FWHM is
smaller than the FP's free spectral range. Scanning the FP's length
we are able to position this narrow transmittance peak to the center of
the interference filter, $l_F$=95.03~$\mu$m, see the
continuous line in the inset in Fig.~\ref{Fig_AFP}.  With this
setting there are no oscillations in the visibility pattern
any more.

From Fig.~\ref{Fig_AFP} we can clearly see that while there is no
interference in the MZ interferometer detuned from the balanced
position by a few millimeters in the idler beam, a high visibility
can be achieved in coincidence-count measurements by placing a narrow
frequency filter in the path of signal photons. This is a direct
consequence of the strong energy correlations between the two
entangled beams. One can also see this experiment from another point of
view: The FP resonator serves as a postselection device that selects
a quasi-monochromatic component of the signal beam. This selection
is done in the idler beam by means of the coincidence-count
measurement, thus yielding a wide autocorrelation function of the
postselected idler beam.


\section{Time correlations}\label{sec_time}

First we describe theoretical results obtained for the Hong-Ou-Mandel
interferometer. The coincidence-count rate $R_n^{HOM}$ as
a function of relative delay $\Delta t$ between the signal and idler photons
is determined in terms of the two-photon amplitude ${A}$
given in Eq.~(\ref{A}) as follows \cite{Perina_PRA59}
\begin{widetext}
\[ R_n^{HOM}(\Delta t) = 1 - \rho^{HOM}(\Delta t),
\mbox{where} \]
\begin{equation}
 \rho^{HOM}(\Delta t) =
   \frac{ \int_{-\infty}^{\infty} dt_A \int_{-\infty}^{\infty} dt_B {\rm Re}
          \left\{ {A}(t_A - \Delta t,t_B) {A}^*(t_B - \Delta t,t_A) \right\}
        }
        { \int_{-\infty}^{\infty} dt_A \int_{-\infty}^{\infty} dt_B
          \left| {A}(t_A,t_B) \right|^2
        }.
 \label{hom-rho}
\end{equation}
\end{widetext}

Assuming that the spectrum of signal and idler photons is determined by
the geometric filtering according to Eq.~(\ref{T_g}) with parameters
$\sigma_1 = \sigma_2 = \sigma$, the interference term $\rho^{HOM}$
of Eq.~(\ref{hom-rho}) assumes the form
\begin{equation}\label{hom-dip}
 \rho^{HOM}(\Delta t) = \exp\left[ - \frac{\sigma^2
 \Delta t^2}{2}
   \right] .
\end{equation}
The width of the dip given by $\rho^{HOM}$ is half the width of
the visibility peak given by Eq.~(\ref{vis_g}) in the case of geometric
filtering without any additional filters.

\begin{figure}[thb]
\centerline{\includegraphics[width=0.75\columnwidth]{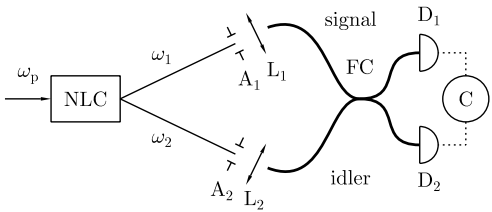}}
\caption{Experimental setup for the Hong-Ou-Mandel interference
measurement.
The photon-pair source (NLC) and the fiber-coupling optics are identical to that 
shown in Fig.~\protect\ref{setup}. FC is a fiber coupler, D$_1$ and D$_2$
are photodetectors, C denotes detection electronics.
} \label{HOM}
\end{figure}

\begin{figure}[hbt]
\centerline{\includegraphics[height=\columnwidth,angle=270]{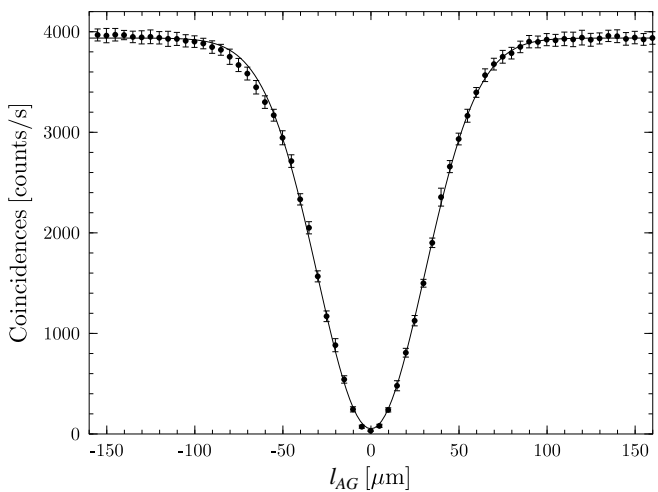}}
\caption{Coincidence-count interference pattern in the Hong-Ou-Mandel
interferometer as a function of the path difference of the interferometer's arms.
The continuous curve is a Gaussian fit given by Eq.~(\protect\ref{hom-dip}). 
The FWHM of the coincidence-count dip is 72~$\mu$m.}
 \label{Fig_dip}
\end{figure}

It is necessary to emphasize that the experiment as performed in the
previous section is not a proof of the strong quantum correlations 
produced by the entangled state of Eq.~(\ref{Psi2}).
For instance, if the source emitted 
two monochromatic photons with classically correlated frequencies 
$\omega_{1}$ and $\omega_{2}$ (in the sense of a statistical ensemble) 
with a proper classical probability density $p(\omega_{1},\omega_{2})$,
the same increase of visibility in the coincidence-count measurement with a FP 
resonator and a MZ interferometer would be obtained \cite{Dusek_CJP46}. 
However, such a source could never give the tight time correlations 
of detection instants observed in the HOM interferometric setup \cite{Hong_PRL59}. 
In order to confirm the validity of the description based on the entangled 
two-photon state, we performed a time correlation measurement in the HOM
interferometer with our source of photon pairs.

Figure.~\ref{HOM} shows a scheme of the HOM interferometric setup. 
The time difference $\Delta t$ is obtained by translating 
the fiber-coupling collimator in one of the HOM interferometer's arms towards 
the nonlinear crystal.
The width of the measured
coincidence-count dip (see Fig.~\ref{Fig_dip}) is 72~$\mu$m
which yields the coherence time of the down-converted
photons of 240~fs and the spectral FWHM of 6.0~nm. This spectral width is
slightly larger than the one obtained with the MZ interferometer
(5.3~nm). The difference of these two results is due to a slight 
readjustment needed after changing the experimental setup (compare 
Figs.~\ref{HOM} and \ref{setup}).


\section{Conclusions}\label{sec_conclusions}

Second- and fourth-order interference experiments testing the
nature of entangled two-photon fields generated in spontaneous
parametric down-conversion have been studied. Energy correlations
were observed by frequency filtering of signal photons, while the 
idler photons propagated through a Mach-Zehnder interferometer. It
has been demonstrated that even if the MZ
interferometer is unbalanced to the extent that no interference
can be observed, interference with a high visibility is restored 
in the coincidence-count detections provided that the signal
photons are transmitted through a narrow frequency filter.
Theoretical results obtained in the presented model for a Gaussian
filter, Fabry-Perot resonator or both filters together are in
good agreement with our experimental results. Tight time
correlations have been observed in the Hong-Ou-Mandel
interferometer.

It should be stressed that both experiments inspecting energy
correlations and time correlations provide complementary
information about the generated photon state. The same results in
each individual experiment can be obtained with classical fields, but
there is no classical model that could explain both of them
together. This shows that the two-photon state generated in the
process of spontaneous parametric down-conversion is properly
described by the \emph{entangled} two-photon state of Eq.~(\ref{Psi2}),
because it is this entangled two-photon state that can explain both
energy and time correlations together.


\section{ACKNOWLEDGEMENTS}

This research was supported by the Ministry of Education of the Czech
Republic under grants LN00A015, CEZ J14/98, RN19982003012, and
by EU grant under QIPC project IST-1999-13071 (QUICOV). 
The authors would like to thank A. Luk\v{s} and
V. Pe\v{r}inov\'a for help with mathematical derivations.

\end{document}